\documentclass[aps,prb,floatfix,reprint,showpacs,10pt]{revtex4-1}

\usepackage{amsfonts}
\usepackage{amssymb}
\usepackage{amsmath}

\usepackage{epsfig}
\usepackage{natbib}

\begin{document}

\title{Anisotropic optical properties of Fe/GaAs(001) nanolayers from first principles}
\author{Sebastian Putz, Martin Gmitra, Jaroslav Fabian}
\affiliation{Institute for Theoretical Physics, University of Regensburg, D-93040 Regensburg, Germany}

\begin{abstract}
We investigate the anisotropy of the optical properties of thin Fe films on GaAs(001) from first-principles calculations. Both intrinsic and magnetization-induced anisotropy are covered by studying the system in the presence of spin-orbit coupling and external magnetic fields. We use the linearized augmented plane wave method, as implemented in the WIEN2k density functional theory code, to show that the $C_{2v}$ symmetric anisotropy of the spin-orbit coupling fields at the Fe/GaAs(001) interface manifests itself in the corresponding anisotropy of the optical conductivity and the polar magneto-optical Kerr effect. While their magnetization-induced anisotropy is negligible, the intrinsic anisotropy of the optical properties is significant and reflects the underlying $C_{2v}$ symmetry of the Fe/GaAs(001) interface. This suggests that the effects of anisotropic spin-orbit coupling fields in experimentally relevant Fe/GaAs(001) slabs can be studied by purely optical means.
\end{abstract}


\maketitle

\section{Introduction}
\label{introduction}

Spintronics is an attempt to generalize conventional electronics by exploiting the electron spin as an additional degree of freedom. The goal of semiconductor spintronics\cite{Zutic2004, Fabian2007} is to design novel nanoelectronic devices\cite{Datta1990, Jonker2003, Zutic2011} that exploit spin effects, and whose fabrication can be seamlessly integrated into the existing infrastructure for CMOS technology. Semiconductor spintronics rests on three pillars: spin injection, spin manipulation, and spin detection. First, a non-equilibrium spin distribution needs to be created by spin injection from a ferromagnet into a semiconductor. Exploiting spin-orbit coupling (SOC) effects in the material, it can then be manipulated by magnetic or electric fields before the resulting spin distribution is eventually detected.

The Fe/GaAs heterostructure was the first system in which room-temperature spin injection from a ferromagnet into a semiconductor was achieved\cite{Hanbicki2002}. It is an ideal model system for spin injection: Iron contributes its high Curie temperature and spin moment, and gallium arsenide its high carrier mobility and its long spin lifetime. Moreover, the lattice mismatch of these materials is rather small\cite{Chambers1986}. That allows for the epitaxial growth of unstrained Fe/GaAs interfaces, whose preparation is cheap and has been demonstrated repeatedly\cite{Waldrop1979, Krebs1987, Wastlbauer2005}.

Although the Fe/GaAs(001) interface quality is known to be limited by various surface reconstructions and possible interdiffusion processes, it is easier to prepare than the Fe/GaAs(110) variant\cite{Li2004, Gruenebohm2009} and is thus the more widely studied system. The interface structure crucially determines the electronic structure of the system, such as the spin-polarization at the Fermi level and the magnetic moments. It was found that arsenic termination of the Fe/GaAs(001) interface limits diffusion processes and favors a flat interface\cite{Ashraf2011, Fleet2013}. Consequently, an As-terminated flat interface was chosen as the basic structure for the model calculations of the present study.

The microscopic structure of the Fe/GaAs(001) interface exhibits $C_{2v}$ symmetry\cite{Fabian2007}, which manifests itself in many properties of the system. For example, the in-plane magneto-crystalline anisotropy of thin Fe layers on GaAs(001) has a dominant uniaxial contribution\cite{Gester1996, Zoelfl1997, Brockmann1999, Bensch2002, Zakeri2006, Bayreuther2012}. Furthermore, Fe/GaAs(001) shows a small but very robust tunneling anisotropic magneto-resistance effect, which was demonstrated by Moser \textit{et al.} in 2007\cite{Moser2007}. The tunneling anisotropic magneto-thermopower and spin-Seebeck effects are similar phenomena that have been predicted for this system\cite{LopezMonis2014, LopezMonis2014B}. All these anisotropic effects can be directly attributed to the $C_{2v}$ symmetry of the effective SOC magnetic field at the Fe/GaAs(001) interface\cite{MatosAbiagueA, MatosAbiagueB, Gmitra2013}, which includes both Bychkov-Rashba\cite{Bychkov1984} and Dresselhaus\cite{Dresselhaus1955} contributions that stem from the structure inversion asymmetry and the bulk inversion asymmetry of the system, respectively.

In the present work we investigate the influence of the anisotropic interface SOC fields on the optical properties of the Fe/GaAs(001) heterostructure by means of density functional theory model calculations. While we find the direction of magnetization of the Fe layer to have a negligible effect on the optical properties, the intrinsic (or crystallographic) optical anisotropy is significant. We show that it is manifest in an anisotropic optical conductivity as well as an anisotropic polar magneto-optical Kerr effect (AP-MOKE). The latter means that the Kerr angles (rotation and ellipticity) at normal incidence of the probing beam depend on its direction of linear polarization, thus reflecting the underlying anisotropy of the optical constants of the material. We find the AP-MOKE of our Fe/GaAs(001) model system to have the same $C_{2v}$ symmetry as the SOC fields at the interface. The anisotropy that we observe in the optical conductivity of our $C_{2v}$ structure is to be contrasted
with the \textit{absence} of anisotropy in the Boltzmann dc conductivity\cite{Trushin2007} of a two-dimensional electron gas with Bychkov-Rashba and Dresselhaus spin-orbit interactions.

In the next section we present the methods used to perform the calculations of this study. It is followed by a discussion of the results and concluding thoughts.

\section{Method}
\label{Method}

\begin{figure}
 \centering
 \includegraphics[width=0.9\linewidth]{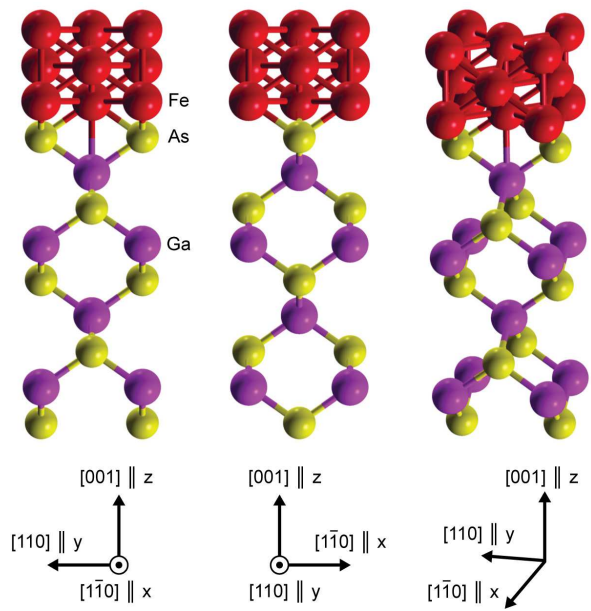}
 \caption{The supercell used to model the Fe/GaAs(001) interface viewed from three different angles. The coordinate tripods indicate the mapping of Cartesian to crystallographic axes.}
 \label{Structure}
\end{figure}

We choose a slab supercell of 15 atoms (see Fig.\,\ref{Structure}) to model an Fe/GaAs(001) interface consisting of 9 monolayers of GaAs(001) and 3 monolayers of Fe. The interface is As-terminated and flat, and the vacuum distance between neighboring slabs is 6\,\AA{}. The structure is not relaxed, but experimental values for the lattice constants and interatomic distances are chosen. Passivation of the structure with hydrogen is not necessary since an explicit calculation shows that the SOC contribution of the As atoms at the bottom of the structure to the optical properties is negligible.

The linearized augmented plane wave method\cite{Andersen1975}, as implemented in the density functional theory (DFT) code package WIEN2k\cite{Blaha2001}, is used to calculate the electronic structure of the system for various directions of the magnetization in the Fe layer. Here, a Monkhorst-Pack\cite{Monkhorst1976} mesh of $(12\times12\times1)$ $k$ points in the full first Brillouin zone is used, and an accuracy of $10^{-7}\,\mathrm{Ry}$ in the total energy is chosen as the convergence criterion. We use the Perdew-Burke-Ernzerhof\cite{Perdew1996} variant of the generalized gradient approximation for the exchange-correlation functional. Spin-orbit coupling is included in all calculations and for all atoms of the supercell using the method of second diagonalization native to WIEN2k. To study the magnetization-induced anisotropy of the system, the calculations are performed for the magnetization $\mathbf{M}$ oriented along $x$, $y$, and $z$ as well as selected intermediate directions in the $xy$ plane. The mapping of Cartesian to crystallographic axes is given in Fig.\,\ref{Structure}.

On top of the converged electronic densities of the previous step we calculate the dielectric function and the optical conductivity in linear response. Here, a much denser Monkhorst-Pack mesh consisting of at least $(70\times70\times1)$ $k$ points is used. The obtained results are converged for all practical purposes. The WIEN2k optics package makes use of the following expression\cite{AmbroschDraxl2006} to calculate the imaginary part of the complex dielectric tensor from the converged Kohn-Sham eigensystem:

\begin{align}
 \mathrm{Im}\left[\epsilon_{\alpha\beta}(\omega)\right] = & \frac{\hbar^2 e^2}{\pi m_e^2 \omega^2} \sum_{n \neq n'} \,\int \! d\mathbf{k} \; \Pi_{n n',\mathbf{k}}^\alpha\,\Pi_{n'n,\mathbf{k}}^\beta\label{DielectricFunction}\\
&\times \left[ f(\epsilon_{n,\mathbf{k}}) - f(\epsilon_{n',\mathbf{k}}) \right] \,\delta(\epsilon_{n',\mathbf{k}} - \epsilon_{n,\mathbf{k}} - \hbar\omega), \nonumber
\label{DielectricTensorFromBandStructure}
\end{align}

where $\Pi_{n n',\mathbf{k}}^\alpha = \langle n',\mathbf{k}|\hat{p}_\alpha | n,\mathbf{k} \rangle$ is the transition matrix element of the $\alpha$ component of the momentum operator for a direct interband transition ($n \neq n'$) from the initial Kohn-Sham state $|n,\mathbf{k}\rangle$ with energy $\epsilon_{n,\mathbf{k}}$ into the final state $|n',\mathbf{k}\rangle$ with energy $\epsilon_{n',\mathbf{k}}$. The Fermi-Dirac distribution function evaluated at energy $\epsilon_{n,\mathbf{k}}$ is given by $f(\epsilon_{n,\mathbf{k}})$, $m_e$ denotes the electron mass, and $\omega$ is the angular frequency of the electromagnetic radiation causing the transition.

The $k$-space integration uses the Bl\"ochl tetrahedron method\cite{Bloechl1994} and only direct interband transitions from occupied to unoccupied bands up to an energy of 20 eV above the Fermi level are taken into account. An energy resolution of 13.6 meV is chosen for the photon energy $\hbar\omega$, and a Lorentzian broadening of 100 meV is applied to account for finite-lifetime effects. Since that broadening distorts the results unphysically for energies on the same order, results for energies smaller than 300 meV are not shown.

The real part of the complex dielectric tensor is obtained from the imaginary part by a Kramers-Kronig transformation. The complex optical conductivity tensor can then be calculated according to\cite{Dressel2001}

\begin{equation}
 \sigma_{\alpha\beta}(\omega) = \frac{\omega}{4\pi i} \left( \epsilon_{\alpha\beta}(\omega) - \delta_{\alpha\beta} \right).
 \label{EpsilonToSigma}
\end{equation}

The AP-MOKE rotation and ellipticity angles depend both on the photon energy $\hbar\omega$ and the polarization state of the probing light beam at normal incidence, which means propagation along the $-z$ direction and linear polarization in the $xy$ plane. Rotating the direction of linear polarization of the incoming beam about the $z$ axis by an angle $\varphi$ is equivalent to rotating the sample by an angle $-\varphi$ in the $xy$ plane. We choose the latter description to simplify the derivation of an expression for AP-MOKE. For the present Fe/GaAs(001) model system, with $\mathbf{M}$ oriented along $z$, the complex dielectric tensor is of the form\cite{Pershan1967, Oppeneer1999, Schoenes1992}

\begin{equation}
 \epsilon = \begin{pmatrix} \epsilon_{xx} & \epsilon_{xy} & 0 \\ -\epsilon_{xy} & \epsilon_{xx} + \delta & 0 \\ 0 & 0 & \epsilon_{zz} \end{pmatrix},
 \label{DielectricTensor}
\end{equation}

where $\delta = \epsilon_{yy} - \epsilon_{xx}$ is a measure of the intrinsic anisotropy in the $xy$ plane. Here and in the following, the explicit dependence of $\epsilon$ and all derived quantities on $\omega$ is suppressed to simplify the notation. A detailed account of the following derivation for a general dielectric tensor can be found in Ref.\,[\onlinecite{Fumagalli1990}].

Let the incoming beam be linearly polarized along the $x$ direction of the fixed coordinate system. The transformed dielectric tensor for the sample rotated by $-\varphi$ about the $z$ axis is then given by

\begin{equation}
 \epsilon' = R_z(-\varphi) \epsilon R_z^T(-\varphi) = \begin{pmatrix} \epsilon'_{xx} & \epsilon'_{xy} & 0 \\ \epsilon'_{yx} & \epsilon'_{yy} & 0 \\ 0 & 0 & \epsilon_{zz} \end{pmatrix},
 \label{TransformedDielectricTensor}
\end{equation}

where the superscript $T$ denotes the transpose. The rotation matrix $R_z(-\varphi)$ is defined as

\begin{equation}
 R_z(-\varphi) = \begin{pmatrix} \cos\varphi & \sin\varphi & 0 \\ -\sin\varphi & \cos\varphi & 0 \\ 0 & 0 & 1 \end{pmatrix}.
\end{equation}

The transformed components of the dielectric tensor are expressed as

\begin{align}
 \epsilon'_{xx} &= \epsilon_{xx} + \frac{1}{2} \big( \delta - \delta \cos(2\varphi) \big), \\
 \epsilon'_{yy} &= \epsilon_{xx} + \frac{1}{2} \big( \delta + \delta \cos(2\varphi) \big), \\
 \epsilon'_{xy} &= \epsilon_{xy} + \delta \cos\varphi \, \sin\varphi, \\
 \epsilon'_{yx} &= -\epsilon_{xy} + \delta \cos\varphi \, \sin\varphi.
\end{align}

We describe the complex electric field vector of polarized light with wave vector $\mathbf{k}$ by the plane wave

\begin{equation}
 \mathbf{E}(\mathbf{r},t) = \mathbf{E}_0 \cdot \exp \left[ i \left( \frac{\omega}{c} \mathbf{n} \cdot \mathbf{r} - \omega t \right) \right],
 \label{PlaneWaveAnsatz}
\end{equation}

where $\mathbf{r}$ and $t$ are the space and time coordinate, respectively, $c$ is the speed of light in vacuum, and $\mathbf{n} = \mathbf{k}c/\omega$ is the vector of complex refractive indices for a wave with wave vector $\mathbf{k}$. In the Jones vector formalism the complex amplitude $\mathbf{E}_0$ is described by its transverse components. For propagation along $-z$ we can thus write

\begin{equation}
 \mathbf{E}_0 = \begin{pmatrix} E_{0,x} \\ E_{0,y} \end{pmatrix}.
\end{equation}

In general, the polarization state of the reflected beam $\mathbf{E}_r$, which has undergone rotation and ellipticity changes, is elliptical. It is determined by the complex reflection matrix $\rho$ acting on the incoming beam $\mathbf{E}_0$:

\begin{equation}
 \mathbf{E}_r = \begin{pmatrix} E_{r,x} \\ E_{r,y} \end{pmatrix} = \begin{pmatrix} \rho_{xx} & \rho_{xy} \\ \rho_{yx} & \rho_{yy} \end{pmatrix} \begin{pmatrix} E_{0,x} \\ E_{0,y} \end{pmatrix}.
\end{equation}

To derive an expression for AP-MOKE, we need to determine $\rho$ with respect to the polarization direction rotation angle $\varphi$. First, we derive the propagation properties of the normal modes $\mathbf{E}_n$ in the material from the transformed complex dielectric tensor $\epsilon'$ using the Fresnel equation

\begin{equation}
 \left( \epsilon' + \mathbf{n}_j \otimes \mathbf{n}_j - n_j^2 \mathbf{1} \right) \cdot \mathbf{E}_{n,j} = 0.
\label{FresnelEquation}
\end{equation}

Here, $\mathbf{1}$ is the $3\times3$ unit matrix, the solution for $\mathbf{E}_{n,j}$ is the $j$-th normal mode of the material (of which there are in general only two), and $n_j = |\mathbf{n}_j|$ is the absolute value of the complex refractive index vector $\mathbf{n}_j$ associated with the normal mode $\mathbf{E}_{n,j}$.

At normal incidence, which means for light propagation along $-z$, the vector of complex refractive indices takes the form $\mathbf{n} = (0,0,n)^\mathrm{T}$. This, combined with Eqs.\,\eqref{TransformedDielectricTensor} and \eqref{FresnelEquation}, yields two solutions $n_j$ ($j = 1,2$) with

\begin{equation}
 n_j^2 = \frac{\epsilon'_{xx} + \epsilon'_{yy}}{2} \pm \sqrt{\frac{\delta^2}{4} - \epsilon_{xy}^{'2}},
\end{equation}

whereas the associated normal modes are described by the parameter $\beta_j$ ($j = 1,2$) given by

\begin{equation}
 \beta_j = \frac{E_{n,j,y}}{E_{n,j,x}} = \frac{1}{2 \epsilon'_{xy}} (\delta \pm \sqrt{\delta^2 - 4 \epsilon_{xy}^{'2}}).
\end{equation}

One can then show that the components of the reflection matrix $\rho$ are given by

\begin{align}
 \rho_{xx} &= \frac{\beta_2 (1-n_1)(1+n_2) - \beta_1(1+n_1)(1-n_2)}{(\beta_2 - \beta_1)(1+n_1)(1+n_2)}, \\
 \rho_{yy} &= \frac{\beta_2 (1+n_1)(1-n_2) - \beta_1(1-n_1)(1+n_2)}{(\beta_2 - \beta_1)(1+n_1)(1+n_2)}, \\
 \rho_{yx} &= \frac{2\beta_1\beta_2(n_2 - n_1)}{(\beta_2 - \beta_1)(1+n_1)(1+n_2)}, \\
 \rho_{xy} &= \frac{-2(n_2 - n_1)}{(\beta_2 - \beta_1)(1+n_1)(1+n_2)}.
\end{align}

Finally, the total rotation $\theta_\mathrm{tot}$ and ellipticity $\varepsilon_\mathrm{tot}$ caused by a material with dielectric tensor $\epsilon$ can be written as

\begin{align}
 \theta_\mathrm{tot} &= \frac{1}{2} \arctan \left( \frac{-2 \, \mathrm{Re}[\rho_{xx}\bar{\rho}_{yx}]}{|\rho_{xx}|^2 - |\rho_{yx}|^2} \right), \\
 \varepsilon_\mathrm{tot} &= \frac{1}{2} \arcsin \left( \frac{-2 \, \mathrm{Im}[\rho_{xx}\bar{\rho}_{yx}]}{|\rho_{xx}|^2 + |\rho_{yx}|^2} \right),
\end{align}

where a horizontal bar indicates complex conjugation. Those quantities include contributions from both the diagonal components of $\epsilon$, which are even in $\mathbf{M}$, and the off-diagonal components, which are odd in $\mathbf{M}$\cite{Oppeneer1992}. Only the latter are of magneto-optical origin and contribute to the pure AP-MOKE, which is why we need to extract their contribution to $\theta_\mathrm{tot}$ and $\varepsilon_\mathrm{tot}$. We can expand those quantities as follows:

\begin{align}
 \theta_\mathrm{tot}(M) &= \theta_K M + \theta_\mathrm{diag} M^2, \\
 \varepsilon_\mathrm{tot}(M) &= \varepsilon_K M + \varepsilon_\mathrm{diag} M^2.
\end{align}

Here, the index $K$ indicates a Kerr effect quantity, ``diag'' refers to the contribution of the diagonal components, and $M$ is the value of the magnetization along $z$. The Kerr rotation and the  Kerr ellipticity are then given by

\begin{align}
 \theta_K &= \frac{1}{2} (\theta_\mathrm{tot}(M) - \theta_\mathrm{tot}(-M)), \\
 \varepsilon_K &= \frac{1}{2} (\varepsilon_\mathrm{tot}(M) - \varepsilon_\mathrm{tot}(-M)).
\end{align}

The dielectric tensor for the $-M$ case is obtained by performing the transformation $\epsilon_{xy} \rightarrow -\epsilon_{xy}$ in Eq.\,\eqref{DielectricTensor}. Combining Eqs.\,(6)--(8), (14), (15), (16), (18), (20), (21), (24), and (25), together with the numerical \textit{ab initio} results for the dielectric tensor (3), allows us to determine the energy-dependent AP-MOKE rotation and ellipticity for an incoming beam at normal incidence whose direction of linear polarization in the $xy$ plane is rotated by $\varphi$ with respect to the $x$ axis.

\section{Results}
\label{Results}

\begin{figure}
 \centering
 \includegraphics[width=0.95\linewidth]{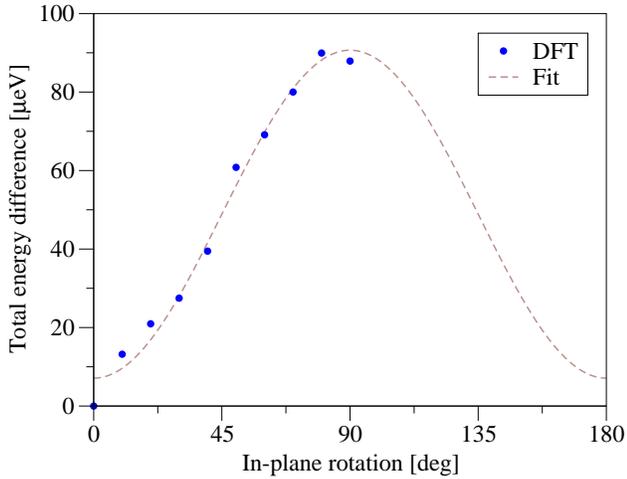}
 \caption{(Color online) Calculated anisotropy of $E_\mathrm{tot}$ (filled blue circles) with respect to the angle between $\mathbf{M}$ in the $xy$ plane and the $x$ axis, relative to $E_\mathrm{tot}^\mathrm{min}$. The DFT results are fitted with a $C_{2v}$ symmetric function (dashed brown line).}
 \label{TotalEnergy}
\end{figure}

In order to study the anisotropy of the model system with respect to the direction of the magnetization $\mathbf{M}$, we calculate the total energy per supercell $E_\mathrm{tot}$ and the real and imaginary part of $\sigma_{xx}$ for $\mathbf{M}$ oriented along $x$, $y$, and $z$ as well as selected intermediate directions in the $xy$ plane. The total energy $E_\mathrm{tot}$ exhibits a $C_{2v}$ symmetric anisotropy of about 88 $\mu$eV (see Fig.\,\ref{TotalEnergy}), with a minimum of $E_\mathrm{tot}^\mathrm{min} = -53437.34893468$ Ry for $\mathbf{M}$ oriented along $x$, and a maximum of $E_\mathrm{tot}^\mathrm{max} = -53437.34892822$ Ry for $\mathbf{M}$ oriented along $y$. In contrast, the anisotropy of the diagonal components of $\sigma$ with respect to the orientation of $\mathbf{M}$ is found to be negligible over the whole calculated energy range (not shown). Note that all subsequent results are obtained for $\mathbf{M}$ oriented along $z$, and that $\sigma$ is given in cgs units.

\subsection{Intrinsic Anisotropy}
\label{IntrinsicAnisotropy}

\begin{figure}
 \centering
 \includegraphics[width=\linewidth]{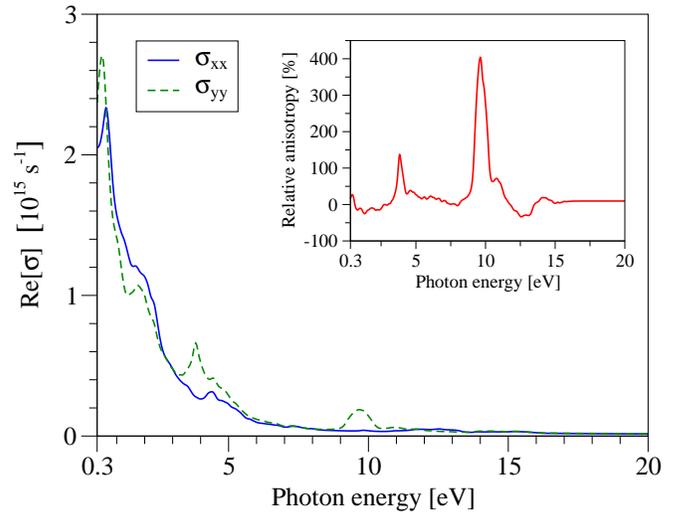}
 \caption{(Color online) Real part of $\sigma_{xx}$ (solid blue line) and $\sigma_{yy}$ (dashed green line) with respect to the photon energy. The inset shows the relative intrinsic anisotropy $A_{xy}^\mathrm{Re}$ (solid red line) with respect to the photon energy.}
 \label{AnisoReal}
\end{figure}

\begin{figure}
 \centering
 \includegraphics[width=\linewidth]{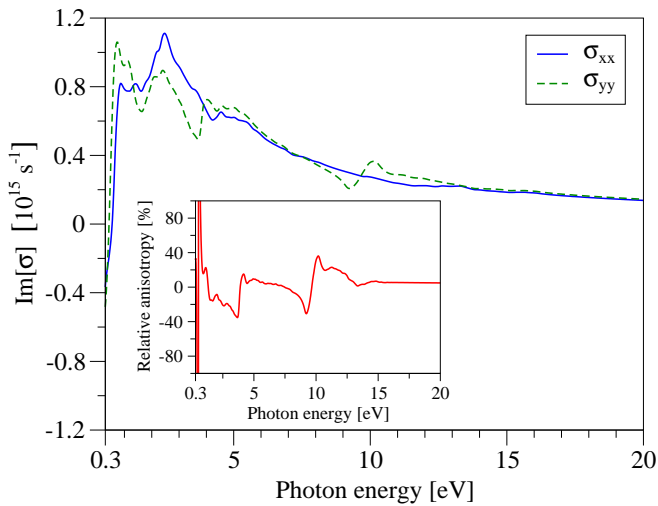}
 \caption{(Color online) See the caption of Fig.\,\ref{AnisoReal}, but for the imaginary part.}
 \label{AnisoImag}
\end{figure}

\begin{figure}
 \centering
 \includegraphics[width=\linewidth]{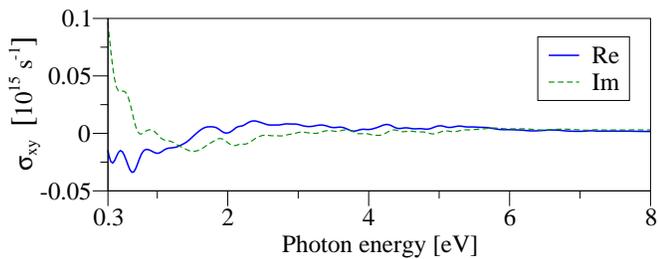}
 \caption{(Color online) Real (solid blue line) and imaginary part (dashed green line) of the off-diagonal component $\sigma_{xy}$ of the optical conductivity with respect to the photon energy.}
 \label{OffDiagonal}
\end{figure}

In contrast to its negligible magnetization-induced anisotropy, the Fe/GaAs(001) model system exhibits a significant intrinsic anisotropy. Figure \ref{AnisoReal} shows the differing real parts of $\sigma_{xx}$ and $\sigma_{yy}$. Both components show pronounced peaks of different height centered at 0.5 eV ($\sigma_{yy}$) and 0.65 eV ($\sigma_{xx}$), and an asymptotic decrease to zero for high energies. The $\sigma_{yy}$ component shows three intermediate peaks where the relative deviation of the two components is largest. The inset of Fig.\,\ref{AnisoReal} presents that relative intrinsic anisotropy $A_{xy}^\mathrm{Re}$ of the real parts of $\sigma_{xx}$ and $\sigma_{yy}$ calculated according to

\begin{equation}
 A_{xy}^\mathrm{Re} = \frac{\mathrm{Re}[\sigma_{yy}] - \mathrm{Re}[\sigma_{xx}]}{\mathrm{Re}[\sigma_{xx}]}.
 \label{eq:CrystAnisoRealRel}
\end{equation}

While $A_{xy}^\mathrm{Re}$ is on the order of tens of percent in the visible range of the electromagnetic spectrum, it reaches values above 100\% and 400\% for 3.9 and 9.6 eV, respectively. Equivalent results for the imaginary part of the optical conductivity are presented in Fig.\,\ref{AnisoImag}. The corresponding relative intrinsic anistropy $A_{xy}^\mathrm{Im}$, given by

\begin{equation}
 A_{xy}^\mathrm{Im} = \frac{\mathrm{Im}[\sigma_{yy}] - \mathrm{Im}[\sigma_{xx}]}{\mathrm{Im}[\sigma_{xx}]},
 \label{eq:CrystAnisoImagRel}
\end{equation}

is shown in the inset of Fig.\,\ref{AnisoImag}. It is on the order of tens of percent for the most part of the calculated energy range, with pronounced extrema at energies of 3.7, 9.3, and 10.2 eV. The divergence at 0.6 eV is a consequence of $\mathrm{Im}[\sigma_{xx}]$ crossing zero at that energy.

Well converged off-diagonal components of the optical conductivity tensor and the dielectric tensor are crucial for the calculation of magneto-optical quantities such as the Kerr rotation $\theta_K$ or the Kerr ellipticity $\varepsilon_K$. Figure \ref{OffDiagonal} shows the result for the off-diagonal component $\sigma_{xy}$, which is found to be sufficiently converged at 8100 $k$ points in the full first Brillouin zone. Note that it is an order of magnitude smaller than the diagonal components.

\subsection{Anisotropic Polar MOKE}
\label{AnisotropicPolarMOKE}

\begin{figure}
 \centering
 \includegraphics[width=\linewidth]{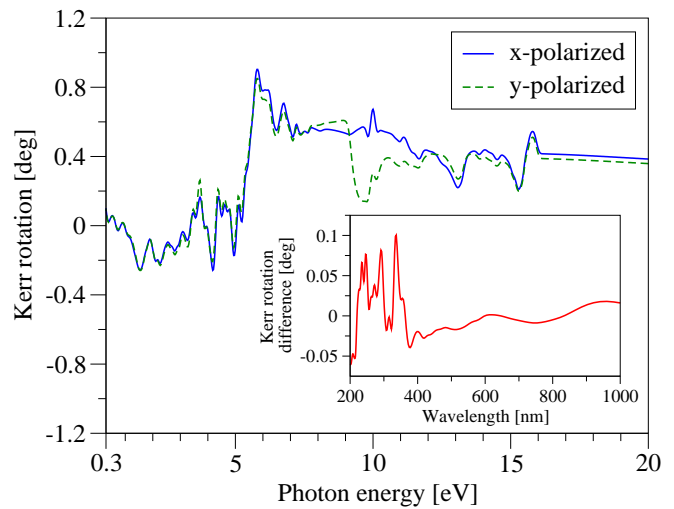}
 \caption{(Color online) Kerr rotation for beams linearly polarized along $x$ (solid blue line) and $y$ (dashed green line) at normal incidence with respect to the photon energy. The inset shows the absolute difference in Kerr rotation (solid red line) for $x$- and $y$-polarized incoming beams with respect to the wavelength.}
 \label{KerrRot}
\end{figure}

\begin{figure}
 \centering
 \includegraphics[width=0.99\linewidth]{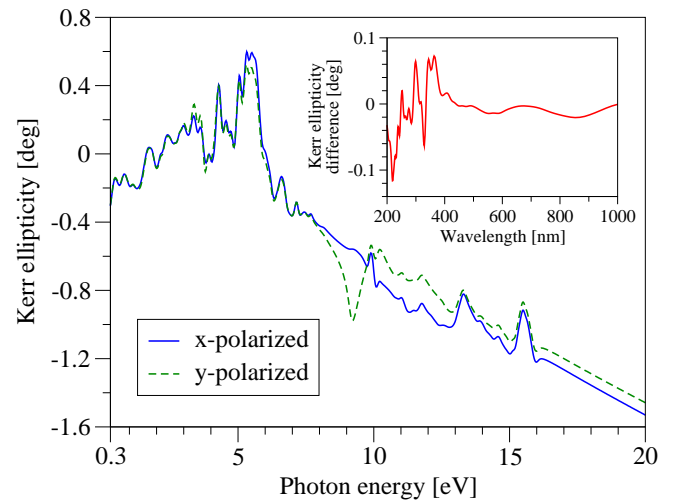}
 \caption{(Color online) See the caption of Fig.\,\ref{KerrRot}, but for the Kerr ellipticity.}
 \label{KerrEllip}
\end{figure}

\begin{figure}
 \centering
 \includegraphics[width=\linewidth]{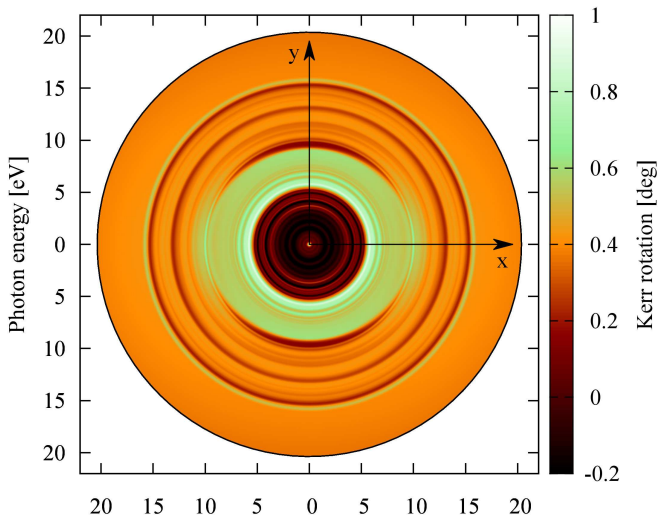}
 \caption{(Color online) Kerr rotation of a probing beam incident along $-z$ with respect to the photon energy. The azimuth indicates the angle between the $x$ axis and the direction of linear polarization of the probing beam.}
 \label{APMOKE01}
\end{figure}

\begin{figure}
 \centering
 \includegraphics[width=\linewidth]{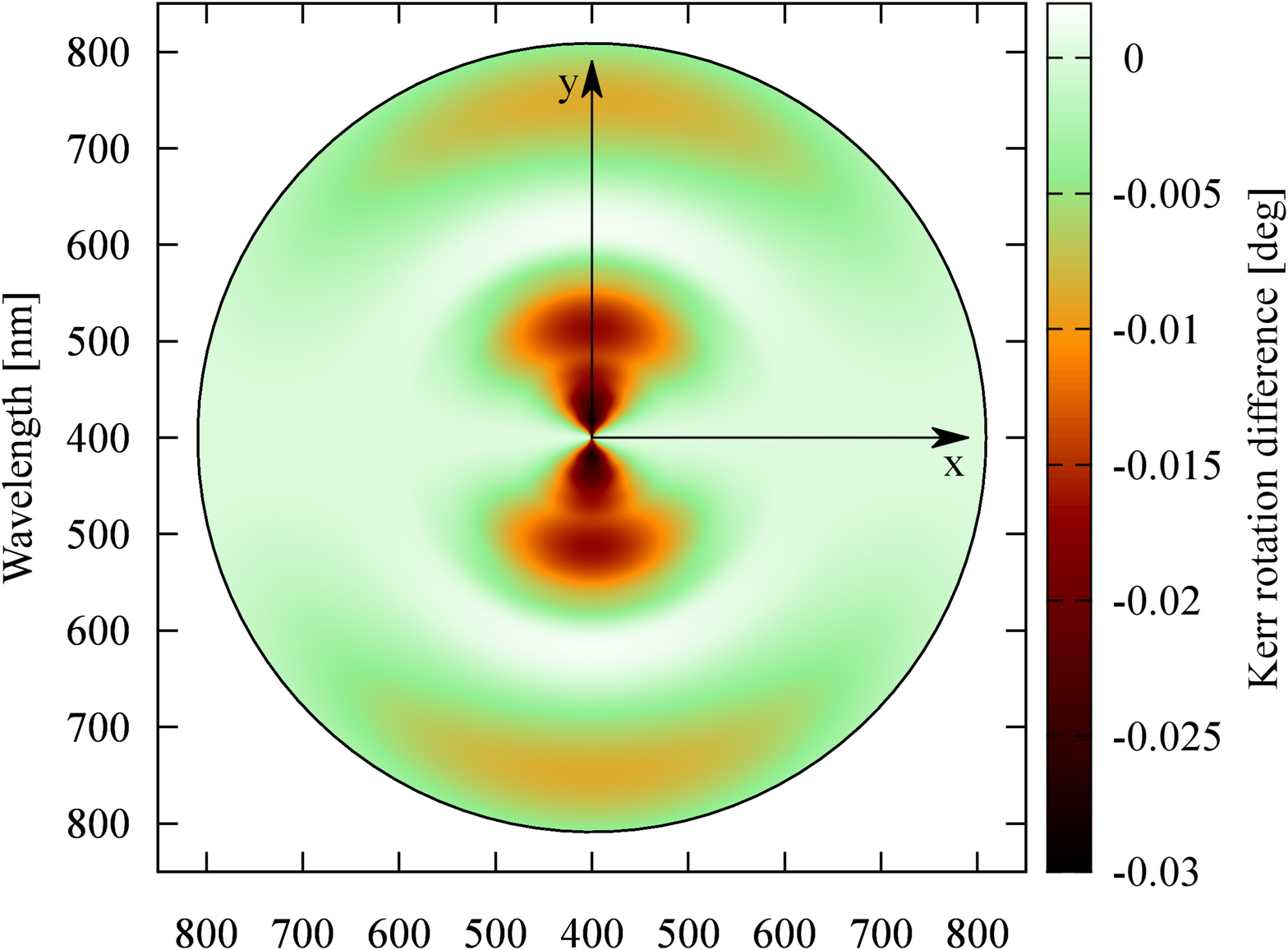}
 \caption{(Color online) Kerr rotation of a probing beam incident along $-z$, relative to the Kerr rotation of a probing beam polarized along the $x$ direction, with respect to the wavelength. The azimuth indicates the angle between the $x$ axis and the direction of linear polarization of the probing beam.}
 \label{APMOKE02}
\end{figure}

\begin{figure}
 \centering
 \includegraphics[width=\linewidth]{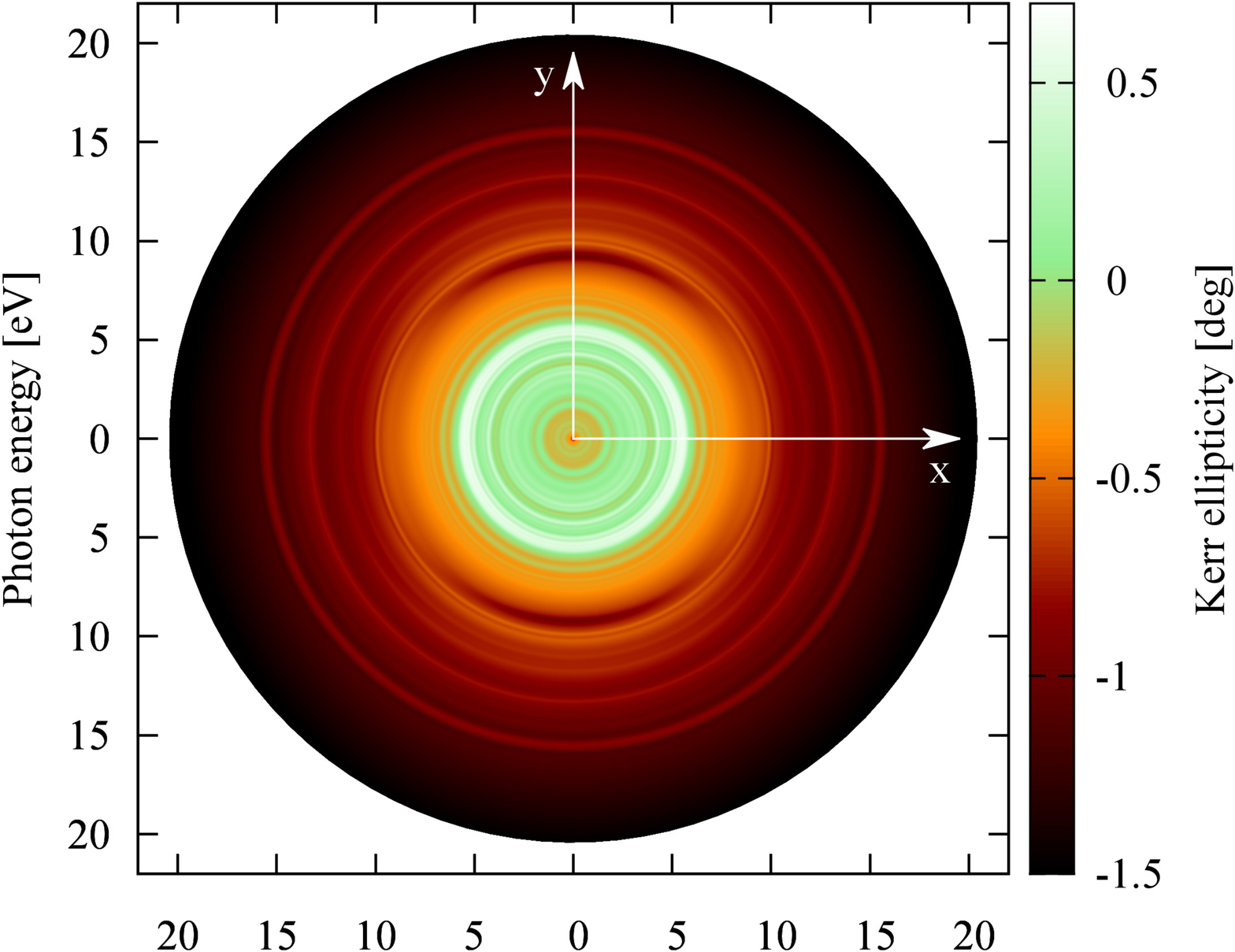}
 \caption{(Color online) See the caption of Fig.\,\ref{APMOKE01}, but for the Kerr ellipticity.}
 \label{APMOKE03}
\end{figure}

\begin{figure}
 \centering
 \includegraphics[width=\linewidth]{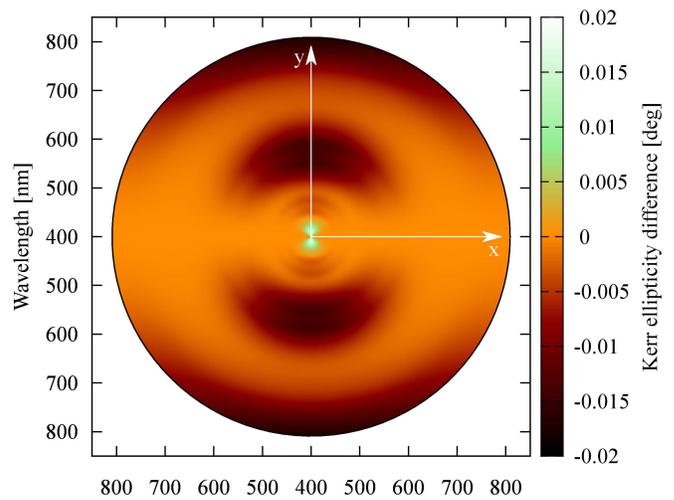}
 \caption{(Color online) See the caption of Fig.\,\ref{APMOKE02}, but for the Kerr ellipticity.}
 \label{APMOKE04}
\end{figure}

The AP-MOKE rotation and ellipticity are obtained according to the procedure described in Sect.\,\ref{Method}. Figure \ref{KerrRot} shows the Kerr rotations for incoming beams polarized along $x$ and $y$ at normal incidence along $-z$. Their absolute values are smaller than $1^\circ$ over the whole calculated energy range, with the largest deviation at an energy of about 10 eV. The inset shows the absolute difference in Kerr rotation for the $x$- and $y$-polarized incoming beams with respect to an experimentally relevant wavelength range. The absolute value of that difference does not exceed $0.1^\circ$ in the given wavelength range.

Analogous results for the Kerr ellipticity are presented in Fig.\,\ref{KerrEllip}. The absolute value of the ellipticity is smaller than $0.5^\circ$ for energies smaller than 8 eV, while it is on the order of $1^\circ$ for higher energies. The largest deviation in Kerr ellipticity for an $x$- and $y$-polarized incoming beam occurs at about 9 eV. The inset shows the absolute difference in Kerr ellipticity for the $x$- and $y$-polarized case with respect to the wavelength. It is bounded by $\pm 0.1^\circ$ over the given wavelength range.

The AP-MOKE for arbitrary linear polarization angles of the incoming beam is illustrated by the polar plots in Figs.\,\ref{APMOKE01}--\ref{APMOKE04}. The azimuth in those plots corresponds to the angle $\varphi$ (see Sect.\,\ref{Method}) between the direction of linear polarization of the incoming beam and the $x$ direction, which corresponds to the crystallographic $[1\bar{1}0]$ direction (see Fig.\,\ref{Structure}). The photon energy or the wavelength are given along the radial direction, and the color scale indicates the respective magneto-optical quantity.

These plots serve to visualize the $C_{2v}$ symmetry of AP-MOKE (with the two mirror axes along $x$ and $y$), which is a manifestation of the underlying effective SOC field symmetry at the Fe/GaAs interface. Depending on its initial polarization state, the reflection of the incoming beam is governed by a transformed dielectric tensor or optical conductivity tensor, which results in an anisotropy of the calculated magneto-optical quantities.

\section{Conclusions}
\label{conclusions}

We studied the anisotropic optical properties of an Fe/GaAs(001) model system from first-principles calculations. While the anisotropy of the optical conductivity with respect to the direction of magnetization in the Fe layer is found to be negligible, the intrinsic anisotropy is significant. The relative intrinsic anisotropy of the real and imaginary part of the optical conductivity in the infrared, visible, and ultraviolet spectrum is on the order of tens of percent, with maxima of up to 100\% and 400\% at certain energies.

In addition to the optical conductivity, the anisotropic polar magneto-optical Kerr effect was studied for arbitrary linear polarization directions of the probing beam at normal incidence. The resulting anisotropic Kerr rotation and Kerr ellipticity reach values up to about $\pm 1^\circ$ and reflect the underlying $C_{2v}$ symmetry of the Fe/GaAs(001) interface.

While interface imperfections and protective capping layers\cite{Tacchi2007, Lei2007} might lead to quantitatively different experimental results, the qualitative results presented here are expected to be observable in high-quality samples using state-of-the-art optical setups.

In conclusion, our results suggest that the effects of the $C_{2v}$ symmetric effective SOC fields at the Fe/GaAs(001) interface can be studied by purely optical means in experimentally relevant samples. Interfacial effects, including lowering of the planar symmetry along the interface, play an increasingly important role in the electronic transport and the optics of nanostructures. By controlling the interface electrically, for example, one could also control the spin-orbit fields and thus modify the electric and optical properties of the connected electronic system (in our case ferromagnetic iron).

\acknowledgments
This work was supported by GRK 1570 and SFB 689 of the German Research Foundation.

\bibliography{FeGaAsPaper}

\end{document}